\documentclass[copyright]{eptcs}
\usepackage{breakurl}
\usepackage{graphicx}
\title{ACL2 Meets the GPU: Formalizing a CUDA-based Parallelizable All-Pairs Shortest Path Algorithm in ACL2}
\author{David S. Hardin
\institute{Advanced Technology Center\\ Rockwell Collins\\
Cedar Rapids, IA, USA}
\email{dshardin@rockwellcollins.com}
\and
Samuel S. Hardin
\institute{Department of Electrical and Computer Engineering\\Iowa State University\\
Ames, IA, USA}
\email{sshardin@iastate.edu}
}
\def\titlerunning{ACL2 Meets the GPU}
\def\authorrunning{D. Hardin \& S. Hardin}
\begin{document}
\maketitle

\begin{abstract}
As Graphics Processing Units (GPUs) have gained in capability and
GPU development environments have matured, developers are 
increasingly turning to the GPU to off-load the main host CPU of  
numerically-intensive, parallelizable computations.  Modern
GPUs feature hundreds of cores, and offer programming niceties such as 
double-precision floating point, and even limited recursion.  This 
shift from CPU to GPU, however, raises the question: how do we know that 
these new GPU-based algorithms are correct?

In order to explore this new verification frontier, we formalized 
a parallelizable all-pairs shortest path (APSP) algorithm for
weighted graphs, originally coded in NVIDIA's CUDA
language, in ACL2.  The ACL2 specification is written using a 
single-threaded object (stobj) and tail recursion, as the stobj/tail
recursion combination yields the most straightforward translation 
from imperative programming languages, as well as efficient, 
scalable executable specifications within ACL2 itself.  The ACL2 
version of the APSP algorithm can process millions of vertices and edges  
with little to no garbage generation, and executes at one-sixth 
the speed of a host-based version of APSP coded in C --- 
a very respectable result for a theorem prover.

In addition to formalizing the APSP algorithm (which uses Dijkstra's
shortest path algorithm at its core), we have also provided capability 
that the original APSP code lacked, namely shortest path recovery.
Path recovery is accomplished using a secondary ACL2 stobj
implementing a LIFO stack, which is proven correct.  To conclude the
experiment, we ported the ACL2 version of the APSP kernels 
back to C, resulting in a less than 5\% slowdown, 
and also performed a partial back-port to CUDA, which, 
surprisingly, yielded a slight performance increase.
\end{abstract}

\section{Introduction}\label{intro}
The computer industry is at a (perhaps temporary) plateau in the race
for core speed, but that race has been replaced with a race for more and 
more cores per die.  Graphics Processing Units (GPUs) are at the
forefront of this new race --- modern GPUs now feature hundreds of
cores.  As GPUs have gained in capability (GPUs now offer programming 
niceties such as double-precision floating point, and even limited
recursion) and GPU development environments have matured, developers
are increasingly turning to the GPU to off-load the main host CPU of 
numerically-intensive, parallelizable computations.   This 
shift from CPU to GPU, however, raises the question: how do we know that 
these new GPU-based algorithms are correct?

The question of correctness has several dimensions, including the absence
of data races, correctness of the CPU/GPU tools and interfaces, even
correctness of the GPU instruction set implementation, all of which 
are interesting questions, but ones which we will not focus on 
here.  Instead, we wish to focus on issues of basic
functional correctness.  In this, there is some hope that automated
verification tools such as ACL2 will be of some use: GPU kernels are
small single-threaded programs, written in a restricted dialect of C
(e.g., CUDA \cite{CUDA} or OpenCL \cite{OpenCL}) and utilizing
fixed-size data structures.  While this programming model does not
match up very well with most verification languages (as noted, for example, in
\cite{Hardin2009a}), it is a fair fit to ACL2 using its
single-threaded objects, or stobjs \cite{STOBJ}.  ACL2 is also
known for producing relatively speedy and scalable executable specifications,
which is valuable for validation of any code that we port from GPU
programming languages, such as CUDA or OpenCL, to the ACL2 environment.

Thus, we embarked on the following experiment: we would manually translate a
CUDA or OpenCL kernel (or kernels) into ACL2, and see how difficult it was to admit
such a kernel into the ACL2 environment, as well as how challenging it would be
to do proofs about the translated code.  We would then validate the
translation by executing the ACL2 version of the kernel, and comparing
output with the original.  After examining how well the ACL2
version performed by running it on very large data sets, we would 
``back-port'' the ACL2 version to see how well the back-ported
version performed.  These experiments were performed starting during
Christmas break of 2012, and continued on nights and weekends through 
January 2013.

\section{The ACL2 Theorem Prover}

We utilize the ACL2 theorem proving system \cite{ACL2book} for much of our high-assurance verification work, as it best presents a single model for formal
analysis and simulation.   
ACL2 provides a highly automated theorem proving environment for machine-checked formal 
analysis, and its logic is an applicative subset of Common Lisp
\cite{CommonLispHyperSpec}.  The fact that ACL2 reasons about 
a real programming language suggests that ACL2 could be an appropriate choice for  
GPU code verification work.  An additional feature of ACL2, single-threaded objects, 
adds to its strength as a vehicle for reasoning about fixed-size data
structures, as will be detailed in the following sections.

\subsection{ACL2 Single-Threaded Objects}\label{stobj}

ACL2 enforces restrictions on the declaration and use of
specially-declared structures called 
single-threaded objects, or stobjs \cite{STOBJ}.  From the perspective
of the ACL2 logic, a stobj is just an ordinary ACL2 object, and can be
reasoned about in the usual way.  Ordinary ACL2 functions are employed to
``access'' and ``update'' stobj fields (defined in terms of the list
operators \texttt{nth} and \texttt{update-nth}).  However, ACL2 enforces strict 
syntactic rules on stobjs to ensure that ``old'' states of a stobj are
guaranteed not to exist.  
This property means that ACL2 can provide destructive implementation for stobjs, 
allowing stobj operations to execute quickly.  In short, an ACL2 single-threaded 
object combines a functional semantics about which we can readily
reason, utilizing ACL2's powerful heuristics,  with a relatively
high-speed imperative implementation that more closely follows 
``normal'' programming practice.

However, there is no free lunch.   As noted by several researchers 
(e.g. \cite{Hardin2009a}), reasoning about functions on stobjs is more difficult
than performing proofs about traditional ACL2 functions on lists which 
utilize cdr recursion.  This difficulty is compounded by the fact
that in order to scale to data structures containing millions of
elements, all of our functions must be tail-recursive (this would be
the case whether we used stobjs or not).  It is also well-known in the
ACL2 community that tail-recursive functions are more difficult to
reason about than their non-tail-recursive counterparts.  One of the
contributions of this work is a preliminary method to deal with these
issues, at least for the case of functions that operate over large
stobj arrays.  With this method, we have been able to show, for a number of such 
functions, that a tail-recursive, stobj-based predicate that marches 
from lower array indices to upper ones is equivalent to a 
cdr-recursive version of that predicate operating over a simple list.
This technique, which we have named Hardin's Bridge\footnote{In memory
  of Scott Hardin, father and grandfather of the two authors: a civil engineer who designed
  several physical bridges, and a man who valued rigor.}, relates a
traditional imperative loop operating
on an array of values to a primitive recursion operating on a
list.

\subsection{A Bridge to Primitive Recursion}\label{bridge}

\begin{figure}
\begin{center}
\includegraphics[scale=0.75]{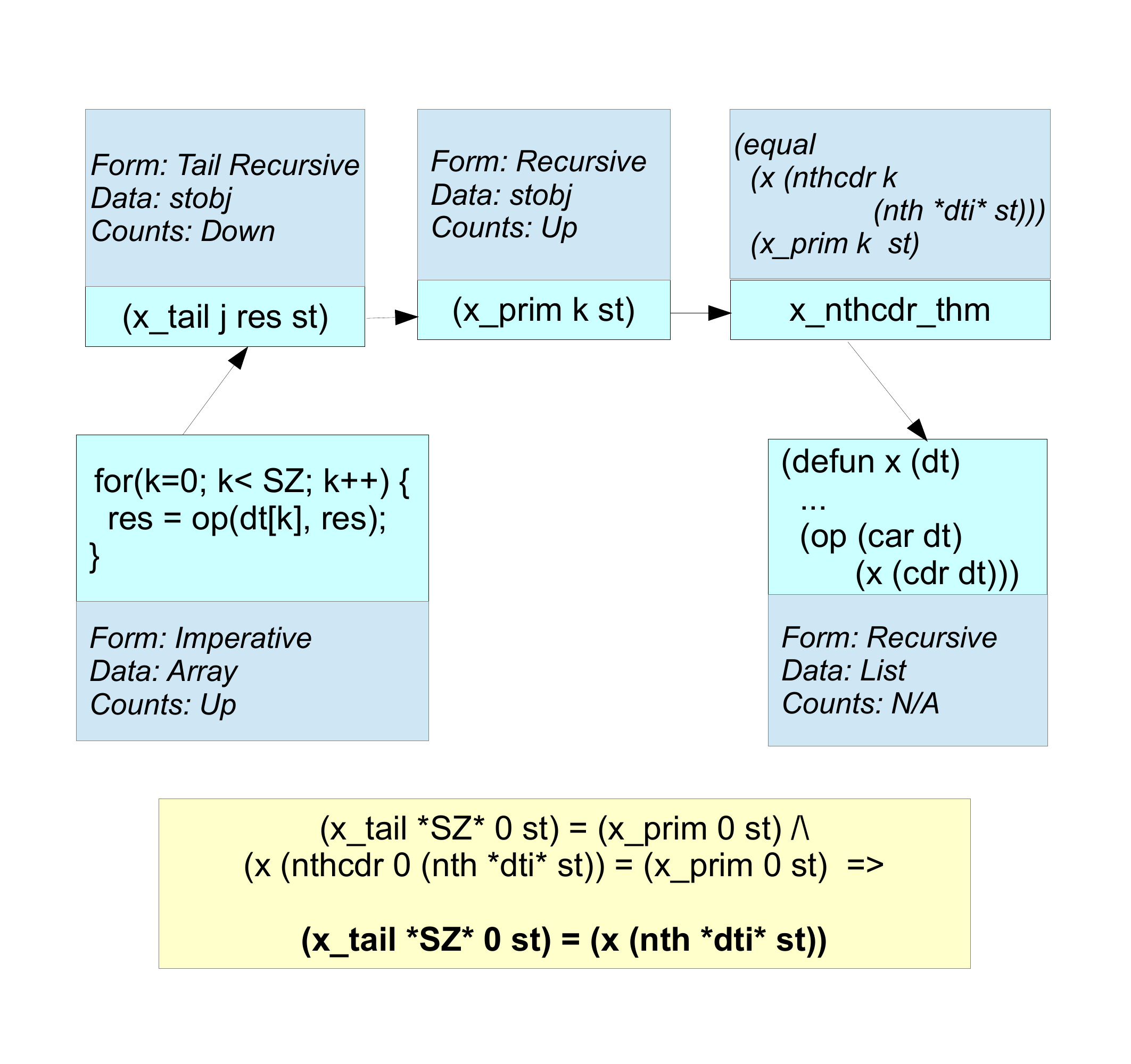}
\end{center}
\caption{Hardin's Bridge: Relating an imperative loop to a primitive-recursive ACL2 function.}
\label{hardins-bridge-pic}
\end{figure}

As depicated in Figure \ref{hardins-bridge-pic}, the process of
building this bridge 
begins by translating an imperative loop, which operates using \texttt{op} on an array
\texttt{dt}, into a tail-recursive function (call it \texttt{x\_tail}) 
operating on a stobj \texttt{st} containing an array
field, also named \texttt{dt}.  This tail-recursive function is invoked as
\texttt{(x\_tail j res st)}, where \texttt{res} is an accumulator, and
the index \texttt{j} counts down from the size of the data array,
\texttt{*SZ*}, toward zero.  However, the data is indexed as
\texttt{(dti (- *SZ* j) st)}, maintaining the same direction of 
``march'' through the data (i.e., from low to high indices) as the 
original, imperative loop.  \texttt{x\_tail} is then shown
to be equivalent to a non-tail-recursive function \texttt{x\_prim}
that also operates over the stobj \texttt{st}.  This function is
invoked as \texttt{(x\_prim k st)}, where index \texttt{k} counts
up from 0 to \texttt{(1- *SZ*)}.  The equivalence of these two
functions is established in a manner similar to the
\texttt{defiteration} macro found in 
\texttt{centaur/misc/iter.lisp} in the distributed books, but with a
change of variable to adjust the count direction.

We now have a primitive recursion that operates on an array field of a
stobj.  What we desire, however, is a primitive recursion that
operates over lists.  One such recursion is as follows:

\begin{verbatim}
(defun x (lst)
  (cond ((endp lst) val)
        (t (op (car lst) (x (cdr lst))))))
\end{verbatim}

This can be related to \texttt{x\_prim} by way of a theorem involving
\texttt{nthcdr}:

\begin{verbatim}
(defthm x_nthcdr_thm
  (implies (and (stp st) (natp k) (< k *SZ*))
           (= (x (nthcdr k (nth *DTI* st))) 
              (x_prim k st))))
\end{verbatim}

By functional instantiation (setting k = 0), if the preconditions are met, then
\begin{verbatim}
(equal (x (nth *DTI* st)) (x_prim 0 st))
\end{verbatim}
and, by the earlier equivalence,
\begin{verbatim}
(equal (x_tail *SZ* 0 st) (x_prim 0 st)
\end{verbatim}
so, finally,
\begin{verbatim}
(equal (x_tail *SZ* 0 st) (x (nth *DTI* st)))
\end{verbatim}
We can now prove theorems about the array-based iteration by
reasoning about \texttt{x}, which has a much more convenient form.  
We have employed this bridge technique on several predicates and
mutators, including an array-based insertion sort employing a nested loop.

\section{GPU Programming: CUDA and OpenCL}

Much GPU programming is conducted in either CUDA or OpenCL, although
there are many other language alternatives, most of which are bindings
to an underlying CUDA or OpenCL system.  CUDA is associated with GPUs
from NVIDIA; OpenCL supports GPUs from a number of vendors.  
Both CUDA and OpenCL are a ``superset of a subset'' of the C family 
of languages; CUDA supports C++ features, while OpenCL is based 
on the C99 standard.  As one might expect, these languages do not 
support many of the more powerful (and, one might add, dangerous) 
features of C, such as function pointers, bit fields, and 
variable-length arrays and structs --- this is actually good news from
the formal verification perspective.  A bit of not-so-good news is
that OpenCL currently does not support recursion, although CUDA does 
(to a limited degree).  Both languages are maturing rapidly, though --- 
double precision floating point is now allowed, for instance, on GPU 
hardware that supports it.

One of the nice extensions that both of these languages provide is the
vector type.  (For reasons of space, when discussing
language details, we will focus on CUDA; OpenCL has many similar 
features.)  CUDA supports \texttt{<type>m} vector types, where
\texttt{<type>} is a scalar type, and m=1, 2, 3, or 4.  
Creating and initializing a vector is accomplished as follows: 

\begin{verbatim}
  float4 f = make_float4(0.0, 1.0, 2.0, 3.0);
\end{verbatim}
Accessing a vector component is accomplished by
the \texttt{.[xyzw]} convention.  For 
example, the zeroth component of an \texttt{int4} variable \texttt{j} 
containing four integers is \texttt{j.x}, and the last component is \texttt{j.w}. 

The GPU is designed to be a coprocessor to a host CPU.  As the 
GPU comprises many cores, most of the extended language 
features of CUDA and OpenCL have to do with setup and parallel 
execution of the cores of this coprocessor.  In the CUDA parallel model, any
core can theoretically access any memory space; however, there is no
guarantee of data coherency.  On the other hand, the callers of and
execution site for a given function can be restricted.  CUDA designates three
function type qualifiers: \texttt{\_\_device\_\_}, \texttt{\_\_global\_\_},
and \texttt{\_\_host\_\_}.  \texttt{\_\_device\_\_} functions may only execute
on the GPU device, and may only be called by other functions on the
device.  \texttt{\_\_global\_\_} declares a function as being a
kernel function, which executes on the GPU device, but is callable from the
host only.  Finally, \texttt{\_\_host\_\_} functions, as the name implies,
execute on the host, and are callable from the host only.

The CUDA hardware model consists of a number of multiprocessors,
which execute asynchronously in parallel.  Each multiprocessor
consists of a number of stream processors, or more commonly, cores.  
Each kernel is dispatched by the host CPU onto a multiprocessor, where it
is divided into groups of threads.  Each thread is part of a block,
and each block is one element of a grid.  All threads in a grid execute
in Single Instruction, Multiple Data (SIMD) fashion.  Each thread
within a block and each block within a grid are given unique
identifiers that can be referenced in the developer's CUDA source code.  
Note that in order to take advantage of potential data parallelism, 
the developer must write her code in a particular way, e.g. array indices need to be 
stratified by block ID and thread ID, as in:

\begin{verbatim}
  int tid = blockIdx.x*MAX_THREADS_PER_BLOCK + threadIdx.x;
\end{verbatim}

\section{The GPU Example: All-Pairs Shortest Path}

In scanning the GPU literature for examples, we looked for
implementations of algorithms that would be familiar to a
theorem proving audience, and that primarily involved integer
operations.  After a bit of searching, we found an interesting 
example in a CUDA implementation of an all-pairs shortest path (APSP) 
algorithm for weighted graphs documented by Harish and Narayanan 
\cite{Harish2007}.  The algorithm was later translated to OpenCL by 
Ginsburg \cite{Ginsburg2011}.  Notably, Ginsburg found an algorithmic error in
Harish and Narayanan's original paper, so we were keen to explore this issue as
well.

\begin{figure}
\begin{center}
\includegraphics[scale=0.5]{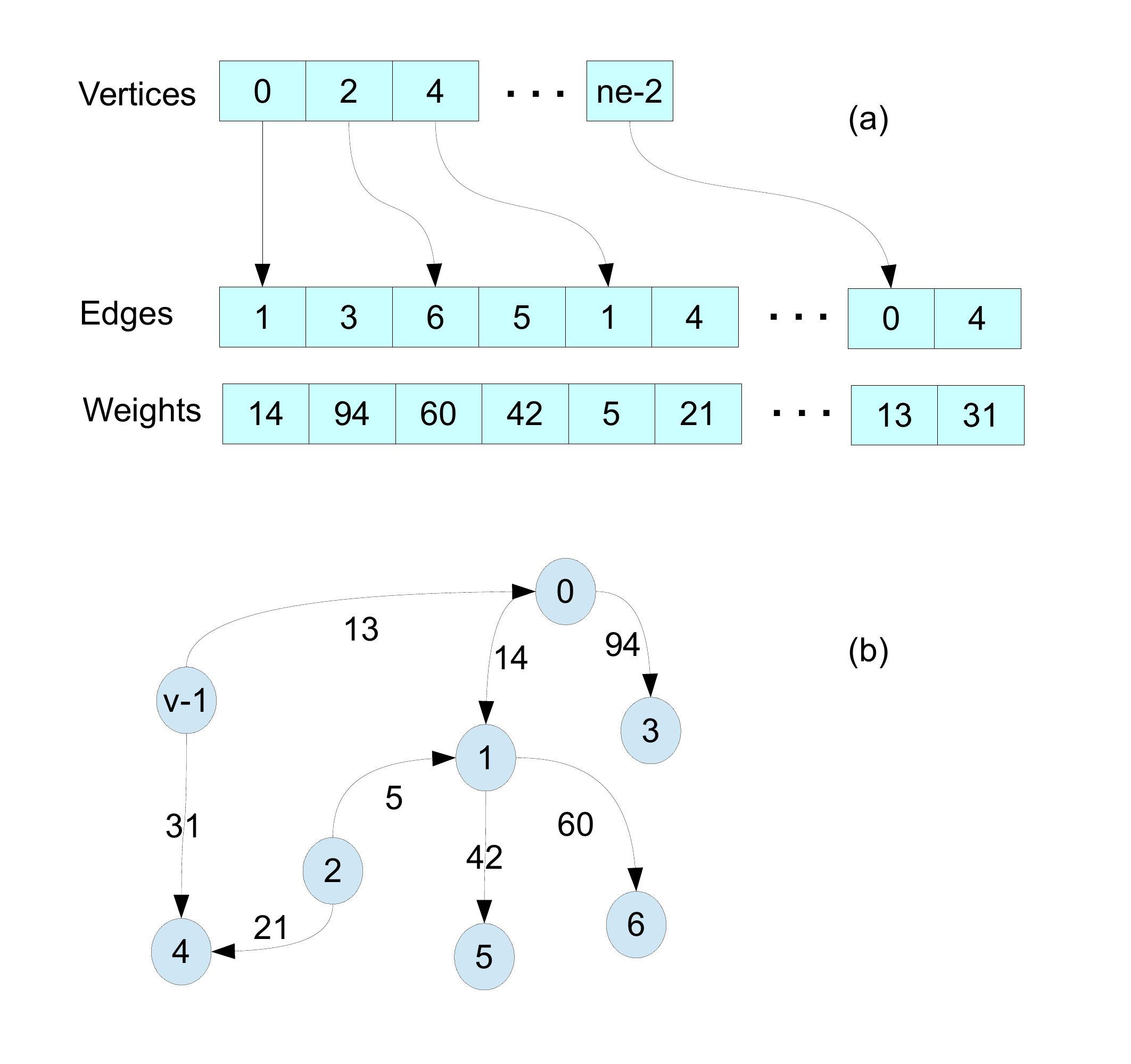}
\end{center}
\caption{(a) Vertices, Edges, and Weight Arrays.  (b) Graph fragment
  from the data in (a).}
\label{vertices-edges-pic}
\end{figure}

The APSP algorithm proceeds by repeated application of Dijkstra's
single-source shortest path (SSSP) algorithm, applied to a data structure in
which the graph vertices, edges, weights, and computed costs are
encoded as arrays.  The vertex array contains indices into the edge
array, whereas the edge array contains vertex indices, as shown in
Figure \ref{vertices-edges-pic}.  The weight array contains the weight
of each edge, and thus is the same size as the edge array.  The cost
array contains the accumulated weight of the shortest path from a
given source vertex to each destination vertex, as computed by
Dijkstra's algorithm.  The costs for the individual shortest path runs
are copied into a final result array of size \texttt{*MAX\_VERTICES* *
  *MAX\_SOURCES*}.  For a full application of APSP, that size would be
the square of the number of vertices.  For large graphs, a full APSP would
exceed the memory capacity of current personal computers, and would take 
quite a long time even if the memory were available.  Thus the 
number of source vertices is usually set to be quite a bit less than the
number of vertices.   (n.b.: If full APSP were desired for a graph of
a million vertices, for example, one could perform the algorithm on,
say, 4000 source vertices at a time, and record the shortest path cost 
data to disk.  Fortunately, this is quite a rare use case.)

One will immediately note a deficiency in this data structure: nowhere
are the shortest paths themselves stored, just their costs.
Apparently, this is common for APSP algorithms, no doubt due to space 
considerations, and also due to the fact that most shortest paths are
not ``interesting'' to the analyst.  Fortunately, once one has the shortest
path cost from vertex A to B, it is relatively easy to recover the shortest
path from A to B.  We have written an ACL2 function to perform path
recovery, using an ACL2 stobj that implements a LIFO stack.

\subsection{Harish's Algorithm}

Harish and Narayanan \cite{Harish2007} present a pseudocode overview
of their APSP implementation in their Algorithm 7.  We reproduce that 
algorithm, which we will refer to from now on as Harish's algorithm, below:

\noindent
\\
Copy vertex array $V_{a}$, edge array $E_{a}$, and weight array $W_{a}$
from G(V, E, W)\\
Create mask array $M_{a}$, cost array $C_{a}$, and updating cost array 
$U_{a}$ of size V\\
\textbf{for} $S$ from 1 to $V$ \textbf{do}\\
\indent$M_{a}$[S] $\gets$ true\\
\indent$C_{a}$[S] $\gets$ 0\\
\indent \textbf{while} $M_{a}$ not Empty \textbf{do}\\
\indent\indent \textbf{for} each vertex V in parallel \textbf{do}\\
\indent\indent\indent Invoke CUDA\_SSSP\_KERNEL1($V_{a}$,$E_{a}$,$W_{a}$,$M_{a}$,$C_{a}$,$U_{a}$) on the grid\\
\indent\indent\indent Invoke CUDA\_SSSP\_KERNEL2($V_{a}$,$E_{a}$,$W_{a}$,$M_{a}$,$C_{a}$,$U_{a}$) on the grid\\
\indent\indent  \textbf{end for}\\
\indent \textbf{end while}\\
\textbf{end for}\\

Harish and Narayanan observe that for large graphs and modern CPU/GPU
architectures, this algorithm performs better than previously 
developed parallel APSP algorithms, such as the parallel Floyd
Warshall APSP algorithm --- the intermediate results are only of size
$O(V)$, and the algorithm involves an $O(V)$ operation looping over 
$O(V)$ threads, as opposed to an $O(V)$ operation looping over 
$O(V^{2})$ threads \cite{Harish2007}.

In Harish's algorithm, the two SSSP kernels proceed as follows in each
major iteration (that is, while the mask array is not all false).  
The first SSSP kernel checks whether the mask entry for each vertex is set.  
If the mask for a given vertex is set, it fetches its current cost from the
cost array, as well as the neighboring edge weights from the weight
array.  The cost of each neighbor vertex is updated if its current
cost is greater than the cost of the current vertex plus the weight 
of the edge from the original vertex to the neighbor vertex.  
This updated cost is recorded not in the cost array, but in a
secondary updating cost array, due to the lack of memory coherence 
in the CUDA programming model.  The second kernel then updates 
each element of the cost array to the value in the updating cost array 
value if the updating cost array value is less, and also sets the 
corresponding mask array value so that the algorithm will examine 
that vertex on a future iteration for further possible cost
reductions.  Otherwise, the updating cost array element is set to 
the value of the corresponding cost array element.

To get a feel for what the kernels look like, let's examine the
first kernel of the APSP algorithm  (Figure \ref{SSSP-kernel1}),
courtesy of Harish \cite{HarishWeb} (with a few cosmetic changes 
added by the authors).

\begin{figure*}
\begin{verbatim}
__global__ void DijkstraKernel1(
    int* g_graph_nodes, int* g_graph_edges, 
    short int* g_graph_weights, int* g_graph_updating_cost, 
    bool* g_graph_mask, int* g_cost, int no_of_nodes, 
    int edge_list_size)
{
  int tid = blockIdx.x*MAX_THREADS_PER_BLOCK + threadIdx.x;
  int i, end, id;
  if (tid < no_of_nodes && g_graph_mask[tid]) {
    if (tid < no_of_nodes-1) {
      end = g_graph_nodes[tid+1];
    } else {
      end = edge_list_size;
    }
    for (i = g_graph_nodes[tid]; i < end; i++) {
      id = g_graph_edges[i];
      atomicMin(&g_graph_updating_cost[id], 
                g_cost[tid] + g_graph_weights[i]);
    }
    g_graph_mask[tid] = false;
  }
}
\end{verbatim}
\hrulefill
\caption{A Single-Source Shortest Path Kernel in CUDA.}
\label{SSSP-kernel1}
\end{figure*}

As noted above, the basic index (\texttt{tid}) is expressed in terms of
the block ID as well as the thread ID within the block; this allows
the CUDA compiler to distribute the data parallel processing to the
cores in a grid.  The other parallel programming wrinkle apparent in
this code is the use of the \texttt{atomicMin} function, which as its
name implies, atomically updates the value pointed to by its first 
argument with the minimum of the value pointed to by the first
argument and the second argument.

\section{APSP in ACL2}

Our translation of Harish's algorithm into ACL2 was aided by the
existence of Ginsburg's OpenCL port \cite{Ginsburg2011}.  Not only did 
Ginsburg point out a bug in the presentation of
\texttt{CUDA\_SSSP\_Kernel1}, but he also provided, in the source code 
accompanying his book chapter, a non-parallel reference implementation 
in C, which was quite helpful.  So, following
Ginsburg's lead, we also decided to develop a non-parallel reference 
implementation, this time in ACL2.  We could possibly have explored 
the use of ACL2(p) \cite{ACL2p} in our modeling, and developed a more 
faithful parallel implementation, but were uncertain about the interplay
of ACL2(p) and stobjs.  (One of the reviewers of this paper noted that, indeed,
\texttt{plet} and stobjs are incompatible.)

The APSP implementation in ACL2 is too lengthy to present in
a paper, but the sources will be made available.  We begin our brief tour of
the ACL2 code with a presentation of the basic stobj used by all functions
in the APSP implementation.

\subsection{Graph Definitions in ACL2}\label{arraysetDef}

The basic single-threaded object declaration for a 
weighted graph of 1 million vertices and 10 edges per vertex is as follows:

\begin{verbatim}
(defconst *MAX_VERTICES* 1000000)
(defconst *MAX_EDGES_PER_VERTEX* 10)
(defconst *MAX_EDGES* (* *MAX_VERTICES* *MAX_EDGES_PER_VERTEX*))
(defconst *MAX_SOURCES* 1)
(defconst *MAX_RESULTS* (* *MAX_VERTICES* *MAX_SOURCES*))

(defstobj GraphData
    ;; Vertex count
    (vertexCount :type (integer 0 1000000) :initially 1000000)

    ;; Edge count
    (edgeCount :type (integer 0 10000000) :initially 10000000)

    ;; (V) Contains a pointer to the edge list for each vertex
    (vertexArray :type (array (integer 0 9999999) (*MAX_VERTICES*)) 
                 :initially 0)

    ;; (E) Contains pointers to the vertices that each edge is 
    ;; attached to
    (edgeArray :type (array (integer 0 999999) (*MAX_EDGES*)) 
               :initially 0)

    ;; (W) Weight array
    (weightArray :type (array (integer 0 *) (*MAX_EDGES*)) 
                 :initially 0)

    ;; (M) Mask array
    (maskArray :type (array (integer 0 1) (*MAX_VERTICES*)) 
               :initially 0)

    ;; (C) Cost array
    (costArray :type (array (integer 0 *) (*MAX_VERTICES*)) 
               :initially 0)

    ;; (U) Updating cost array
    (updatingCostArray :type (array (integer 0 *) (*MAX_VERTICES*)) 
                       :initially 0)

    ;; (S) Source Vertices
    (SourceVertexArray :type (array (integer 0 99) (*MAX_SOURCES*)) 
                       :initially 0)

    ;; (R) Results
    (ResultArray :type (array (integer 0 *) (*MAX_RESULTS*)) 
                 :initially 0))
\end{verbatim}

First note that the algorithm is fundamentally parameterized by the constants
\texttt{*MAX\_VERTICES*}, \texttt{*MAX\_EDGES\_PER\_VERTEX*}, and
\texttt{*MAX\_SOURCES*}.  The latter can be adjusted up to
\texttt{*MAX\_VERTICES*}, in the case of a true all-pairs shortest path
computation.  Next, observe that first two elements of the stobj,
\texttt{vertexCount} and \texttt{edgeCount}, are not strictly
necessary.  But, they were present in Ginsburg's OpenCL and
(non-parallel) C reference versions, so we duplicate them here.  The 
presence of these two elements at the beginning of the stobj 
also stops ACL2 from rewriting, e.g. 
\texttt{(nth *EDGEARRAYI*  GraphData)} into 
\texttt{(cadr GraphData)} during proofs, which 
can be annoying to deal with.  Also observe that we are using arbitrary precision
integers for most data here.  Happily, we have been able to proceed at
reasonable speed without resorting to, e.g. 64-bit integers.  And
remarkably, due to the limited arithmetic involved, we can do
significant runs over large graphs and allocate less than 2000 bytes
of heap data (according to \texttt{time\$}).  Finally,
one may have thought it more elegant to declare the \texttt{maskArray}
elements to be of type \texttt{(or t nil)}.  This turned out to be
both harder to reason about (ACL2 has much better support for 
\texttt{integer-listp}), and significantly slower to execute, so
we abandoned it.

\subsection{ACL2 Translation Techniques}

Once the main data structure was in place, we recoded all loops 
as tail-recursive functions, accepting the \texttt{GraphData} stobj as
a parameter, and returning it if any of its fields were updated.  
Termination proofs were not difficult in general; only one function 
required a dummy ``countdown'' parameter to help with termination analysis.  

One issue that we had in the translation was the treatment of the
various \texttt{for} loops.  In ACL2, it is advantageous to code in a
``decrementing'' style; however, in most imperative coding cultures, 
a \texttt{for} loop that increments an index through an array is the
preferred idiom.  We were somewhat concerned that by changing the 
direction from incrementing to decrementing, that we might
accidentally ``break'' the code by, e.g. failing to notice some data 
dependency of ``later'' elements of an array on ``earlier'' elements.  
The solution we arrived at was to have a decrementing index (call it 
\texttt{ix}), but to address the arrays in question, as, for example, 
\texttt{(vertexArrayi (- *MAX\_VERTICES* ix) GraphData)}.

The result, while somewhat verbose (n.b.: the verbosity could be
hidden with a simple macro), allows the ACL2 code to ``march''
through the array from low indices to high indices as in the original, 
while maintaining a decrementing loop counter parameter that ACL2 
prefers (and whose initial value, conveniently, would be something 
simple like \texttt{*MAX\_VERTICES*}), and also allowing a nice 
\texttt{(zp ix)} test to stop the loop.  So, let us look at the ACL2 
translation of the inner loop of Figure \ref{SSSP-kernel1}, shown in 
Figure \ref{SSSP-kernel1-inner-loop-ACL2}.  The first dozen lines or
so are just necessary guard conditions, and the corresponding checks
in the body of the ACL2 function.  From there, it's a fairly direct
translation from the CUDA inner loop to a tail-recursive function, 
with J Moore's \texttt{seq} macro eliminating some of the \texttt{let} 
binding clutter.

\begin{figure*}
\begin{verbatim}
(defun dsk1-inner-loop (edge-down edge-max tid GraphData)
  (declare (xargs :stobjs GraphData 
                  :guard (and (natp edge-down)
                              (natp edge-max)
                              (natp tid)
                              (< tid *MAX_VERTICES*)
                              (<= edge-down edge-max)
                              (<= edge-max *MAX_EDGES*))))
  (cond ((not (GraphDatap GraphData)) GraphData)
        ((not (natp edge-down)) GraphData)
        ((not (natp edge-max)) GraphData)
        ((not (natp tid)) GraphData)
        ((not (< tid *MAX_VERTICES*)) GraphData)
        ((> edge-down edge-max) GraphData)
        ((> edge-max *MAX_EDGES*) GraphData)
        ((zp edge-down) GraphData)
        (t (let ((updating-index 
                   (edgeArrayi (- edge-max edge-down) GraphData)))
             (seq GraphData 
               (update-updatingCostArrayi 
                 updating-index 
                 (min (updatingCostArrayi updating-index GraphData) 
                      (+ (costArrayi tid GraphData)
                         (weightArrayi 
                           (- edge-max edge-down) GraphData))) 
                 GraphData)
               (dsk1-inner-loop (1- edge-down) edge-max tid 
                                GraphData))))))
\end{verbatim}
\hrulefill
\caption{Single-Source Shortest Path Kernel inner loop in ACL2.}
\label{SSSP-kernel1-inner-loop-ACL2}
\end{figure*}

The entirety of the translation, including all the guard proofs and
related theorems, as well as the occasional comment, comes to nearly 
2600 lines of ACL2 code.  Some of that code, however, provides
functionality not found in the original CUDA code, namely shortest
path recovery.  Some 160 lines of that additional code constitutes 
a stobj-based implementation of a LIFO stack (and its basic 
correctness proofs) that was needed for the shortest path recovery 
implementation.

\subsection{Lessons Learned from the Translation}

The translation to ACL2 revealed several flaws in the original code.
First, as is the case with most C-based code, plain \texttt{int} types were
used when some variant of \texttt{unsigned} types could have been used 
instead.  Since all of the \texttt{GraphData} components are natural numbers, 
and all operations on the \texttt{GraphData} are either additions or 
comparisons, it would have been smarter to use unsigned types.
Additionally, no particular care was taken to guard against
array index overflows, another common problem with C-based code. 
The formalization in ACL2 demonstrates that if the developers had 
spent just a little extra time in declaring and guarding the array
indices, the algorithm could be shown to safely stay within its
various array bounds.

Kernel 1, as originally written, also suffers from a possible integer 
overflow error in the computation of the updating cost array --- 
the addition of the cost array element and the weight element 
could wrap around.  The wrapped-around value could then 
propagate to the updated cost array, as the cost is suddenly 
much lower.  This, in turn, could lead to an inaccurate shortest 
path computation.  Ginsburg's code 
\cite{Ginsburg2011} ameliorates this issue by changing the cost,
weight, and updating cost arrays to use floating point; but this has a
significant impact on performance.  Harish's code
\cite{HarishWeb} makes wrap-around somewhat less likely by declaring
the weight array to be a short (16-bit) integer; nonetheless, in the
worst case one could still have wrap-around of a signed 32-bit cost after 
65,539 additions, which is certainly possibly for shortest path
searches through large graphs.  One could remedy this issue by
changing the min computation in Kernel 1 to something like

\begin{verbatim} 
  cwsum = C[i] + W[j];
  min(U[k], (cwsum < C[i])?WT_MAX:cwsum);
\end{verbatim}

where \texttt{WT\_MAX} is the largest possible weight value.
Preliminary results indicate that the addition of such a wrap-around
check has negligible impact on the algorithm's runtime for large graphs.

The final flaw to be discussed is the error in Kernel 1 disclosed by
Ginsburg.  The problem in the Harish and Narayanan paper
\cite{Harish2007}  (a problem which is 
fixed in the source code retrieved from Harish's web site \cite{HarishWeb}) 
was that the weight array was being indexed by an improper value; 
in essence, it was being indexed 
by an element of the edge array, \texttt{E[i]} instead of just
\texttt{i}.  One can see the problem by noting that weight array
indices should be in the range \texttt{0..*MAX\_EDGES*-1}, 
while the edge array, being an array of vertex indices, should have 
values in the range \texttt{0..*MAX\_VERTICES*-1}, usually a much 
smaller number.  When this error is introduced into 
the ACL2 code, the guard proofs produce a subgoal 
(which it fails to prove without further assistance) whose conclusion 
is basically \texttt{E[i] < *MAX\_EDGES*}.  An analyst with some
knowledge of the data structure would spot this as a strange 
subgoal to generate (as opposed to \texttt{E[i] < *MAX\_VERTICES*}).  
However, this subgoal can be proven eventually, so it can't be said 
that ACL2 found the error.  The only way for ACL2 to uncover the flaw 
would be to attempt an equivalence proof between Kernel 1 and an 
existing version of Dijkstra's algorithm specified in ACL2, a task 
which remains as future work.

\subsection{Guards and Performance}

Given that we wished to validate the ACL2 implementation of APSP
utilizing large graphs (1 million vertices and 10 million edges), we 
needed a reasonably efficient executable specification.  In addition
to exploiting tail recursion and the in-place update capability that
stobjs provide, we also knew that we needed to verify the guards 
of all performance-relevant functions in order to achieve acceptable 
results.  At first, this seemed a daunting task; but it turned out 
to be almost routine after a while.  Many of the guard conjectures we
needed to prove for non-leaf functions stated that if that function 
called another function foo with a \texttt{GraphDatap} argument, 
foo would return a \texttt{GraphDatap} result.  This, in turn, could
be broken down into a series of proofs that the ``types'' of the
components of the \texttt{GraphData} stobj were preserved across a 
call of foo (by ``type'', we mean a basic predicate such as
\texttt{natp}).  Happily, most \texttt{GraphData} components were
unaffected by a given function, and proving that fact was 
particularly easy.  For the other components that were 
updated by function foo, ACL2 usually had little difficulty in
establishing that the updates preserved that component's
``type''.  In the end, having verified guards reduced the execution
time of our tests by 40\%.

One interesting property that arose during guard proof development 
was the need to establish that the \texttt{vertexArray} elements were 
nondecreasing.  (This happens to be the case, given how the vertex
array is initialized by the \texttt{init-vertex-array} function.)
Thus, a number of functions ended up with an additional 
\texttt{(vertices-nondecreasing GraphData)} guard condition.  
Note that we reason about this predicate using the bridge technique 
described in Section \ref{bridge}.

\subsection{Performance Results}

The ACL2 version of Harish's algorithm can find the shortest path 
from a given source vertex to each of a million destination vertices, 
following 10 million edges in less than 8 seconds on a late 2012 
MacBook Pro with a 2.5 GHz Intel Core i5 dual-core CPU, and with 
little to no garbage generation.  The performance of the algorithm is linear 
in the number of source vertices, at least up to the 100 source
vertices tested so far (for a one million vertex graph with 10 edges 
per vertex).

\begin{figure}
\begin{center}
\includegraphics[scale=0.9]{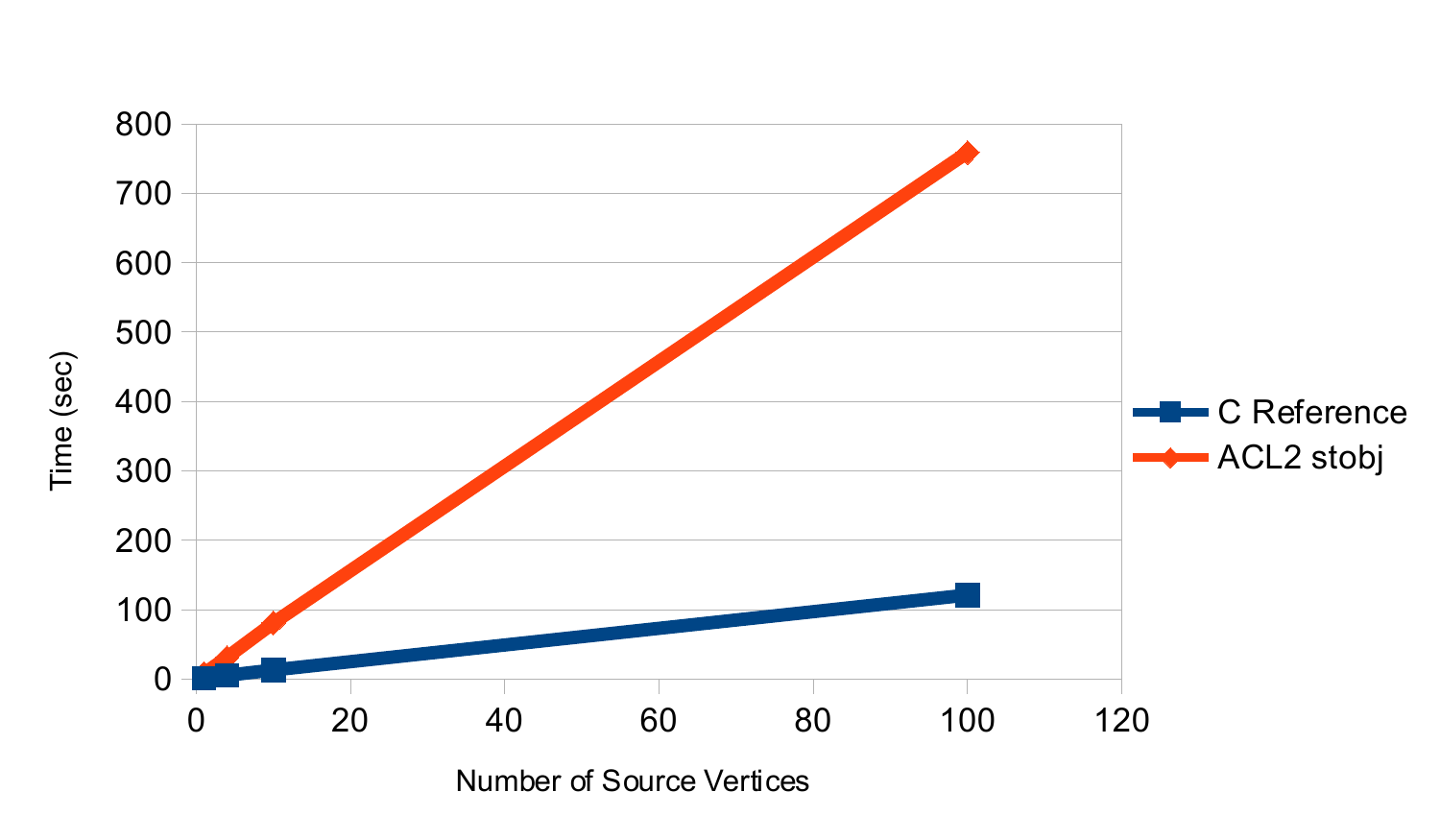}
\end{center}
\caption{Time to find all shortest paths from a number of source vertices.}
\label{APSP-results-pic}
\end{figure}

The performance of the ACL2 version of Harish's algorithm
relative to Ginsburg's non-parallel C reference implementation
is shown as a function of the number of source vertices in Figure 
\ref{APSP-results-pic}.  The performance of both is linear in the
number of source vertices, and the ACL2 version consistently 
executes at one-sixth the speed of the C reference implementation --- 
a very respectable result for a theorem prover, utilizing
arbitrary-precision arithmetic.  As a final comparison, Harish and 
Narayanan report that their GPU-based implementation is 
some 70 times faster than the CPU-only version \cite{Harish2007}.

\section{Back-porting to C and CUDA}

To conclude our experiment, we ported the ACL2 version of the APSP kernels 
back to C and CUDA, to see what effect the tail-recursive style would
have on performance.  We were particularly interested to see if the CUDA
compiler (actually a complicated frontend to GCC) would truly support
tail recursion, and whether the C version would suffer in performance from
moving inner loops into their own functions.  The methodology used 
here was simple, and purely empirical --- keep increasing the 
optimization level on the compiler until performance stopped 
improving or the generated code became broken.  We found that for C 
(using \texttt{clang}), performance of the tail-recursive extracted
inner loops was much poorer than the baseline C reference 
implementation at low optimization levels, but that higher 
optimization (\texttt{-O3}) improved matters significantly, with the 
tail-recursive version showing only a 3\% slowdown.  

In order to obtain results for the back-port of the
tail-recursive ACL2 formulation to CUDA, we first updated Harish's 
code \cite{HarishWeb} to allow it to compile and run in the current
CUDA environment (CUDA 5.0), which we downloaded from the CUDA 
website \cite{CUDA} onto a late 2012 iMac equipped with an 
NVIDIA GeForce GT 650M GPU with 512MB of memory.  
For our initial CUDA back-port, we replaced only the inner loop of 
\texttt{CUDA\_SSSP\_KERNEL1} (see Figure \ref{SSSP-kernel1}) with a 
tail-recursive function hand-translated back from the ACL2 code of Figure 
\ref{SSSP-kernel1-inner-loop-ACL2}.  We conducted complete 
APSP runs (i.e., we computed the shortest paths between all pairs of
vertices) for a randomly generated graph of 100,000 vertices and 5
edges per vertex using both Harish's non-recursive CUDA code 
as well as our tail-recursive alternative.  We timed our results using 
(updated) timing code already present in Harish's sources, and compared the 
output shortest path costs to ensure that the non-recursive 
and tail-recursive versions both yielded the same results.  Surprisingly,
our tail-recursive back-port from ACL2 was slightly faster than
Harish's non-recursive version, but only by 1.8\%.

\section{Related Work}

The use of ACL2 single-threaded objects to speed graph search problems
enjoys a long history.  Wilding \cite{Wilding00} was the
first, in 2000, to use a stobj to speed an ACL2 algorithm, originally
documented by Moore \cite{mooregraph}, for finding a
path between two given vertices in an unweighted graph.  
Like our work, Wilding uses a vertex array and a mask array (perhaps more
descriptively, Wilding calls this array a mark array).  Unlike our 
representation, Wilding's vertex array elements are lists of 
children of that vertex.  Greve and Wilding followed this work 
in 2003 \cite{WildingGreve03} with a version that utilized MBE to 
justify a major optimization used in the 2000 paper.

GPU programming is a relatively new activity, so there has not been 
much research into the verification of GPU programs, toolchains, 
or hardware.  Two notable tools are PUG \cite{PUG}  and GPUVerify
\cite{GPUVerify}.  PUG is a data race analyzer for  CUDA programs, 
utilizing an SMT solver.  GPUVerify translates OpenCL and CUDA 
kernels into Boogie \cite{Boogie} in order to find data races.  
Neither tool focuses on basic functional correctness.

\section{Conclusion and Future Work}

We have formalized a parallelizable all-pairs shortest path (APSP) 
algorithm for weighted graphs, originally coded in NVIDIA's CUDA
language, in ACL2.  The ACL2 specification is written using a 
single-threaded object (stobj) in tail recursive style.  The ACL2 
version of the APSP algorithm scales to millions of vertices and edges  
with little to no garbage generation, and executes at one-sixth 
the speed of a host-based version of APSP coded in C.  We also 
provided capability that the original APSP code lacked, namely 
shortest path recovery.  Path recovery is accomplished using a 
secondary ACL2 stobj implementing a LIFO stack, which was proven
correct.  To bring our work back to where it began, we ported 
the ACL2 version of the APSP kernels back to C, where we 
observed a mere 3\% slowdown, and also performed a partial 
back-port to CUDA, where we measured a slight 
performance increase of 1.8\%.

Future work should focus on using ACL2 to verify that 
Harish's algorithm does indeed implement APSP.  This will 
involve creating an ACL2 formalization of APSP, building 
upon an existing formalization of Dijkstra's shortest path algorithm by 
Moore and Zhang \cite{Dijkstra-SSSP-ACL2}.  Work should also 
continue on improving techniques for reasoning about stobjs; 
some progress has been made during the current effort, but much 
more work is needed in order to make the process easier.

Additional future work on the APSP algorithm implementation 
may involve setting an upper bound 
on the values of elements of the cost, weight, and updating 
cost arrays; as well as investigating the use of
\texttt{unsigned-bytep} types for all arrays.  In the general 
area of GPU verification, many opportunities present 
themselves, including hardware and toolchain 
verification, as well as verification of the CPU/GPU 
interface.

\section{Acknowledgments}

We thank Dave Greve and Konrad Slind for their sage advice, as well as  
the anonymous reviewers for their thorough and thoughtful comments.

\bibliographystyle{eptcs}
\bibliography{fm}
\end{document}